\documentstyle[a4,11pt,multicol,rotating] {article}
\begin{document}

\setlength{\baselineskip}{16pt}

\begin{center}
{\bf\Large Detection of TiH$_2$ molecule in the interstellar medium is less probable}

\vspace{3mm}

{\bf\large Suresh Chandra$^{1,2*}$ \& Mohit K. Sharma$^3$}
\end{center}

\vspace{2mm}

\noindent
$^1$Physics Department, Lovely Professional University, Phagwara 144411, Punjab, India

\noindent
$^2$Zentrum f\"ur Astronomie und Astrophysik, Technische Universit\"at
Berlin, Hardenbergstrasse 36, D-10623 Berlin, Germany

\noindent
$^3$School of Studies in Physics, Jiwaji University, Gwalior 474011, M.P., India

\noindent
Emails: suresh492000@yahoo.co.in; mohitkumarsharma32@yahoo.in

\vspace{4mm}



{\bf Identification of TiH$^1$ and TiO$^2$ has been historical, as the Titanium was first 
time discovered in the interstellar medium (ISM). After finding TiO$_2$$^3$, there is an 
obvious question about the search of titanium dihydride (TiH$_2$). The existence of TiH$_2$
 in the ISM is quite probable, as the atomic abundance of hydrogen is about 1900 times 
larger than that of oxygen. We have discussed that the detection of TiH$_2$ in the ISM is 
less probable, though it has a large electric dipole moment.
}

\vspace{4mm}

For analyzing spectrum from the ISM, one comńsiders an appropriate number of energy levels 
of the molecule of interest. For getting the values of energies of levels and the radiative
 transition probabilities among the levels, one depends on the laboratory study of the 
molecule. To the best of our knowledge, no information about the spectroscopic study of 
TiH$_2$ is available in the literature.

We have optimized the molecule TiH$_2$ with the help of GAUSSIAN 2009, where B3LYP method 
and cc-pVtz basis set are used, and obtained the values of rotational and centrifugal 
distortion constants (Table 1), and of the electric dipole moment. The TiH$_2$ is 
asymmetric top molecule with the Ray parameter $\kappa = -0.6164$. It is a planar molecule 
having C$_{2v}$ symmetry and electric dipole moment $\mu$ = 2.792 Debye along the $b$-axis 
of inertia. The bond length Ti-H is $\sim$1.7077 \AA \ and the angle H-Ti-H is 
$\sim$111.9$^\circ$. 

Owing to one-half value of spin of each of the two hydrogen atoms, the TiH$_2$ has two 
distinct species: (i) ortho (with parallel spins) and (ii) para  (with anti-parallel 
spins). These two species behave as if they are two distinct molecules, as there are no
transitions between them. Since the kinetic temperature in a cosmic object, where molecules
 are found, is few tens of Kelvin, we are concerned with the rotational levels in the 
ground vibrational and ground electronic states. 

\begin{figure}[h]
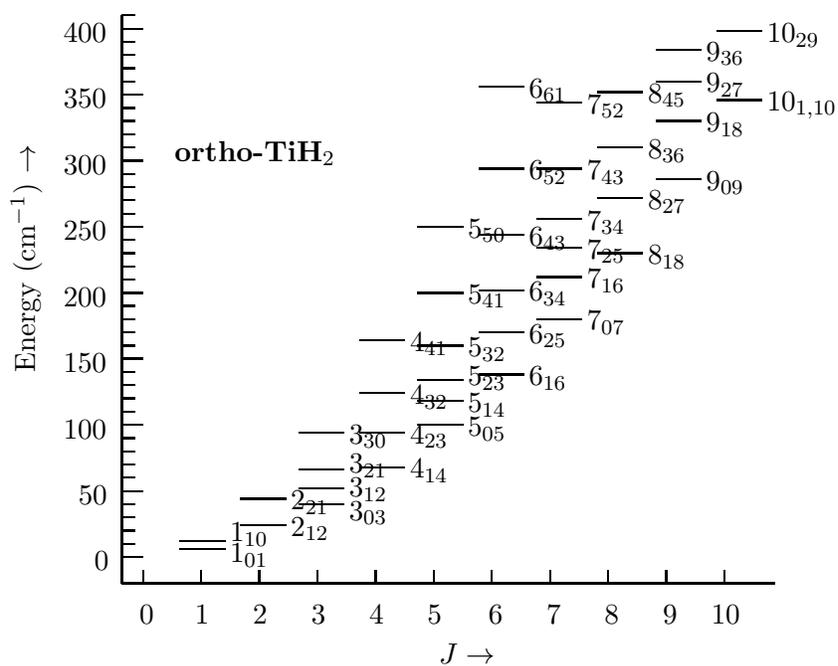

\vspace*{10mm}
\caption{Energy level diagram for 40 levels of ortho-TiH$_2$.}  \label{Fig1A}
\end{figure}

\begin{figure}[h]
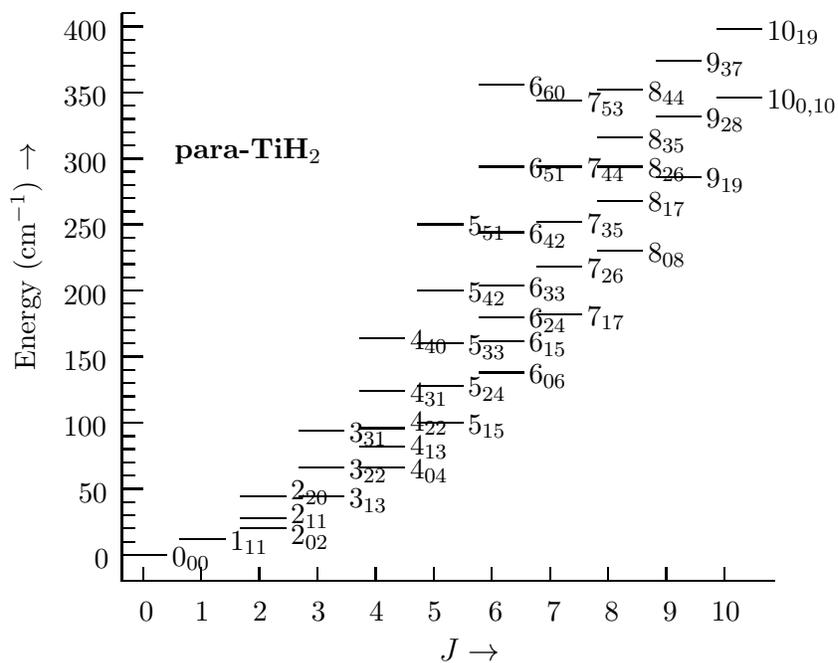

\vspace*{10mm}
\caption{Energy level diagram for 40 levels of para-TiH$_2$.} \label{Fig1B}
\end{figure}

Using the values of rotational and centrifugal distortion constants, we have obtained 
energies of rotational levels. Fot each of the ortho-TiH$_2$ and para-TiH$_2$, 40 
rotational levels are shown in Figures \ref{Fig1A} and \ref{Fig1B}, respectively. These 
levels are connected through radiative and collisional transitions. 

Using the values of rotational and centrifugal distortion constants, and of electric dipole
 moment, the radiative transition probabilities (Einstein $A$-coefficients) are calculated.
 The radiative transitions are governed by some selection rules, derived on the basis of
 quantum mechanics. There are 89 radiative transitions among the ortho levels and 88 
transitions among the para levels. The values of Einstein $A$-coefficients are rather large
 as compared to those of a molecule, in general. For example, out the above transitions, 85
ortho and 83 para transitions have Einstein $A$-coefficient larger than $10^{-5}$ 
s$^{-1}$. The Einstein $A$-coefficient get value up to 1.1 s$^{-1}$.

Though the collisional transitions are not governed by any selection rules, but the
 computation of collisional rate coefficients is the most difficult task in the study of 
cosmic molecules. The collision partner is generally taken as the most abundant molecular
hydrogen H$_2$. When the collisional rate coefficients are not available, it is a 
common practice to use some scaling law for their calculation. We have calculated the
collisional rate coefficients by using a scaling law$^4$. They satisfy the condition that 
they do not produce any kind of anomalous phenomenon from their own. The collisional rate 
coefficients are of the order of 10$^{-11}$ cm$^3$ s$^{-1}$ or smaller. They are, in 
general, smaller than 10$^{-11}$ cm$^3$ s$^{-1}$. 

For the known values of radiative and collisional transition probabilities, for each 
species, we have solved the set of statistical equilibrium equations coupled with the 
equations of radiative transfer. In the model$^5$, the external radiation field, impinging 
on the volume element, generating the lines, is the cosmic microwave background (CMB) only,
 which corresponds to the background temperature $T_{bg}$ = 2.73 K. The parameter 
$\gamma$ is expressed as $\gamma = n_{mol}/(\mbox{d}v_r/\mbox{d}r)$, where $n_{mol}$ is the
 density of TiH$_2$ and $(\mbox{d}v_r/\mbox{d}r)$ the velocity gradient in the cosmic 
object.

The set of equations is non-linear and is solved through iterative procedure, where the
initial population densities of the levels are taken as the thermal ones. In order to 
include a large number of cosmic objects, where titanium dihydride may exist, we have 
considered wide ranges of physical parameters. The density of molecular hydrogen, 
$n_{H_2}$, is taken from 10$^2$ to 10$^7$ cm$^{-3}$, the kinetic temperature $T$ 
is taken as 10, 20, 30, 40, 60, 80, 100 K. Two values of the parameter $\gamma$ are taken 
as $5 \times 10^{-7}$ and $5 \times 10^{-6}$ cm$^{-3}$ (km/s)$^{-1}$ pc. The abundance of 
TiH$_2$ in some cosmic object is expected to be large enough and these values of $\gamma$ 
 are quite reasonable.
 
For the kinetic temperatures considered here, the reliable results may be for the 
transitions among the levels up to $\sim$ 100 cm$^{-1}$. For these levels, there are 14 
ortho and 16 para transitions. The brightness temperatures $T_B$ of the lines are
 calculated by using the population densities of levels obtained by solving the set of 
equations. Except two transitions, $1_{11} - 0_{00}$ and $2_{02} - 1_{11}$ (both belonging 
to the para species), other 28 transitions are found to have low brightness temperature 
$T_B$, which is neaarly equal to the temperatutre of the CMB (2.73 K). The brightness 
temperatures of the transitions $1_{11} - 0_{00}$ and  $2_{02} - 1_{11}$ are shown in 
Figure \ref{Fig2}. For low densities, the brightness temperature is equal to the CMB 
temperature. There is no formation of spectral line. 

For the formation of a spectral line, there should be non-local thermal equilibrium (NLTE)
 in the object. For the NLTE, the collisional transition probabilities are comparable to 
the radiaive transition probabilities. In Figure \ref{Fig2}, the NLTE starts around the 
density of $10^5$ cm$^{-3}$. This value of density is quite large as compared to that found
 ($\sim$ $10^4$ cm$^{-3}$) in a cosmic object having molecules. It may be interpretted that
 the probability of the detection of TiH$_2$ in the ISM is low, whereas the probability of 
its formation is quite large.

It may finally be concluded that owing to the large values of Einstein $A$-coefficients,
the detection of TiH$_2$ in the ISM is less probable.

\vspace{8mm}

\begin{description}

\item{} 1.  Clegg, R.E.S.,  Lambert, D.L. \& Bell, R.A., Astrophys. J. {\bf 234},
188 (1979). 

\item{} 2. Yerle, R.,  Astron. Astrophys. {\bf 73}, 346 (1979).  

\item{} 3. Kaminski, T., Gottlieb, C.A., Menten, K.M.,  {\it et al.}, Astron. Astrophys. 
{\bf 551}, A113 (2013); Kaminski, T., Gottlieb, C.A., Young, K.H., {\it et al.}, Astrophys.
 J. Suppl.  {\bf 209}, 38 (2013); De Beck, E., Vlemmings, W., Muller, S., {\it et al.}, 
Astron.  Astrophys. manuscript no. 25990\_DeBeck, June 3, 2015; arXiv:1506.00818v1 
[astro-ph. SR].

\item{} 4. Sharma, A.K. \& Chandra, S., Astron. Astrophys. {\bf 376}, 333 (2001); Chandra, S., Astron. Astrophys. {\bf 402}, 1 (2003).

\item{} 5. Rausch, E., Kegle, W.H., Tsuji, T. \& Piehler, G., Astron. Astrophys. {\bf 315},
 533 (1996).
\end{description}

\vspace{6mm}

\noindent
{\bf Author Contributions:} S.C. designed the research and wrote the paper. M.K.S. reviewed
 the literature and performed the calculations.

\vspace{6mm}

\noindent
{\bf Author Information:} Correspondence and requests for material should be addressed to 
S.C. (suresh492000@yahoo.co.in).
  

\vspace{26mm}

\begin{tabular}{lr|lr}
\multicolumn{4}{l}{Table 1. Rotational and centrifugal distortion constants in MHz.} \\
\hline
\multicolumn{1}{c}{Constant} & \multicolumn{1}{c|}{Value} & \multicolumn{1}{c}
{Constant} & \multicolumn{1}{c}{Value} \\
\hline
$A$ & $2.8589602 \times 10^{5}$ & $\Phi_{J}$  & $2.037108491 \times 10^{-3}$ \\
$B$ & $1.2520818 \times 10^{5}$ & $\Phi_{JK}$ & $-1.928681064 \times 10^{-2}$  \\
$C$ & $8.707408 \times 10^{4}$ & $\Phi_{KJ}$ & $-1.946479675 \times 10^{-2}$ \\
$\Delta_{J}$ & $6.004343468 $ & $\Phi_{K}$  & $5.110638046  \times 10^{-1}$ \\
$\Delta_{JK}$ & $-4.186548023 \times 10^{1}$ & $\phi_{J}$ & $1.014000418 \times
10^{-3}$ \\
$\Delta_{K}$ & $2.549058853 \times 10^{2}$ & $\phi_{JK}$  & $-3.557138924 \times
10^{-3}$ \\
$\delta_{J}$ & $2.465123434$ & $\phi_{K}$ & $1.085768474 \times 10^{-1}$ \\
$\delta_{K}$ & $2.332384340 $ &   &  \\
\hline
\end{tabular}

\newpage

\vspace*{8.5cm}

 \hspace*{1.5cm}
 \begin{picture}(0,0)
 \put(50,150){\bf ortho-TiH$_2$}
\put( 30, -10){\line(1,0){247}}
\put( 30, -10){\line(0,1){215}}
 \put( 150, -40){$J \rightarrow$}

\put(33, -25){ 0}
\put(54, -25){ 1}
\put(76, -25){ 2}
\put(98, -25){ 3}
\put(120, -25){ 4}
\put(142, -25){ 5}
\put(165, -25){ 6}
\put(187, -25){ 7}
\put(209, -25){ 8}
\put(231, -25){ 9}
\put(253, -25){10}
\multiput(38,-10)(22,0){11}{\line(0,1){5}}
\multiput(30,0)(0,   5){ 42}{\line(1,0){5}}
\multiput(30,0)(0,  25){  9}{\line(1,0){8}}
 \put(-12,70){\rotatebox{90}{Energy (cm$^{-1}$) $\rightarrow$}}

\put(16, -5){  0}
\put(10, 20){ 50}
\put(8, 45){100}
\put(8, 70){150}
\put(8, 95){200}
\put(8,120){250}
\put(8,145){300}
\put(8,170){350}
\put(8,195){400}

\put(  52,   3){\line(1,0){17}}
\put(  71,  -2){$ 1_{0 1}$}
\put(  52,   6){\line(1,0){17}}
\put(  71,   6){$ 1_{1 0}$}
\put(  75,  12){\line(1,0){17}}
\put(  94,   8){$ 2_{1 2}$}
\put(  97,  20){\line(1,0){17}}
\put( 116,  14){$ 3_{0 3}$}
\put(  75,  22){\line(1,0){17}}
\put(  94,  18){$ 2_{2 1}$}
\put(  97,  26){\line(1,0){17}}
\put( 116,  23){$ 3_{1 2}$}
\put(  97,  33){\line(1,0){17}}
\put( 116,  32){$ 3_{2 1}$}
\put( 120,  34){\line(1,0){17}}
\put( 139,  30){$ 4_{1 4}$}
\put( 120,  47){\line(1,0){17}}
\put( 139,  43){$ 4_{2 3}$}
\put(  97,  47){\line(1,0){17}}
\put( 116,  43){$ 3_{3 0}$}
\put( 142,  50){\line(1,0){17}}
\put( 161,  46){$ 5_{0 5}$}
\put( 142,  59){\line(1,0){17}}
\put( 161,  55){$ 5_{1 4}$}
\put( 120,  62){\line(1,0){17}}
\put( 139,  58){$ 4_{3 2}$}
\put( 142,  67){\line(1,0){17}}
\put( 161,  65){$ 5_{2 3}$}
\put( 165,  69){\line(1,0){17}}
\put( 184,  65){$ 6_{1 6}$}
\put( 142,  80){\line(1,0){17}}
\put( 161,  76){$ 5_{3 2}$}
\put( 120,  82){\line(1,0){17}}
\put( 139,  78){$ 4_{4 1}$}
\put( 165,  85){\line(1,0){17}}
\put( 184,  81){$ 6_{2 5}$}
\put( 187,  90){\line(1,0){17}}
\put( 206,  86){$ 7_{0 7}$}
\put( 142, 100){\line(1,0){17}}
\put( 161,  96){$ 5_{4 1}$}
\put( 165, 101){\line(1,0){17}}
\put( 184,  97){$ 6_{3 4}$}
\put( 187, 106){\line(1,0){17}}
\put( 206, 102){$ 7_{1 6}$}
\put( 210, 115){\line(1,0){17}}
\put( 229, 111){$ 8_{1 8}$}
\put( 187, 117){\line(1,0){17}}
\put( 206, 113){$ 7_{2 5}$}
\put( 165, 122){\line(1,0){17}}
\put( 184, 118){$ 6_{4 3}$}
\put( 142, 125){\line(1,0){17}}
\put( 161, 121){$ 5_{5 0}$}
\put( 187, 128){\line(1,0){17}}
\put( 206, 124){$ 7_{3 4}$}
\put( 210, 136){\line(1,0){17}}
\put( 229, 132){$ 8_{2 7}$}
\put( 232, 143){\line(1,0){17}}
\put( 251, 139){$ 9_{0 9}$}
\put( 187, 147){\line(1,0){17}}
\put( 206, 143){$ 7_{4 3}$}
\put( 165, 147){\line(1,0){17}}
\put( 184, 143){$ 6_{5 2}$}
\put( 210, 155){\line(1,0){17}}
\put( 229, 151){$ 8_{3 6}$}
\put( 232, 165){\line(1,0){17}}
\put( 251, 161){$ 9_{1 8}$}
\put( 187, 172){\line(1,0){17}}
\put( 206, 168){$ 7_{5 2}$}
\put( 255, 173){\line(1,0){17}}
\put( 274, 169){$10_{1,10}$}
\put( 210, 176){\line(1,0){17}}
\put( 229, 172){$ 8_{4 5}$}
\put( 165, 178){\line(1,0){17}}
\put( 184, 174){$ 6_{6 1}$}
\put( 232, 180){\line(1,0){17}}
\put( 251, 176){$ 9_{2 7}$}
\put( 232, 192){\line(1,0){17}}
\put( 251, 188){$ 9_{3 6}$}
\put( 255, 199){\line(1,0){17}}
\put( 274, 195){$10_{2 9}$}
 \end{picture}

 \vspace*{8.5cm}

 \hspace*{1.5cm}
 \begin{picture}(0,0)
 \put(50,150){\bf para-TiH$_2$}
\put( 30, -10){\line(1,0){247}}
\put( 30, -10){\line(0,1){215}}
 \put( 150, -40){\large $J \rightarrow$}

\put(33, -25){ 0}
\put(54, -25){ 1}
\put(76, -25){ 2}
\put(98, -25){ 3}
\put(120, -25){ 4}
\put(142, -25){ 5}
\put(165, -25){ 6}
\put(187, -25){ 7}
\put(209, -25){ 8}
\put(231, -25){ 9}
\put(253, -25){10}
\multiput(38,-10)(22,0){11}{\line(0,1){5}}
\multiput(30,0)(0,   5){ 42}{\line(1,0){5}}
\multiput(30,0)(0,  25){  9}{\line(1,0){8}}
 \put(-12,70){\rotatebox{90}{Energy (cm$^{-1}$) $\rightarrow$}}

\put(16, -5){  0}
\put(10, 20){ 50}
\put(8, 45){100}
\put(8, 70){150}
\put(8, 95){200}
\put(8,120){250}
\put(8,145){300}
\put(8,170){350}
\put(8,195){400}

\put(  30,   0){\line(1,0){17}}
\put(  49,  -4){$ 0_{0 0}$}
\put(  52,   6){\line(1,0){17}}
\put(  71,   2){$ 1_{1 1}$}
\put(  75,  10){\line(1,0){17}}
\put(  94,   4){$ 2_{0 2}$}
\put(  75,  14){\line(1,0){17}}
\put(  94,  12){$ 2_{1 1}$}
\put(  97,  22){\line(1,0){17}}
\put( 116,  18){$ 3_{1 3}$}
\put(  75,  22){\line(1,0){17}}
\put(  94,  21){$ 2_{2 0}$}
\put(  97,  33){\line(1,0){17}}
\put( 116,  29){$ 3_{2 2}$}
\put( 120,  33){\line(1,0){17}}
\put( 139,  29){$ 4_{0 4}$}
\put( 120,  41){\line(1,0){17}}
\put( 139,  38){$ 4_{1 3}$}
\put(  97,  47){\line(1,0){17}}
\put( 116,  43){$ 3_{3 1}$}
\put( 120,  48){\line(1,0){17}}
\put( 139,  47){$ 4_{2 2}$}
\put( 142,  50){\line(1,0){17}}
\put( 161,  46){$ 5_{1 5}$}
\put( 120,  62){\line(1,0){17}}
\put( 139,  58){$ 4_{3 1}$}
\put( 142,  64){\line(1,0){17}}
\put( 161,  60){$ 5_{2 4}$}
\put( 165,  69){\line(1,0){17}}
\put( 184,  65){$ 6_{0 6}$}
\put( 142,  80){\line(1,0){17}}
\put( 161,  76){$ 5_{3 3}$}
\put( 165,  81){\line(1,0){17}}
\put( 184,  77){$ 6_{1 5}$}
\put( 120,  82){\line(1,0){17}}
\put( 139,  78){$ 4_{4 0}$}
\put( 165,  90){\line(1,0){17}}
\put( 184,  86){$ 6_{2 4}$}
\put( 187,  91){\line(1,0){17}}
\put( 206,  87){$ 7_{1 7}$}
\put( 142, 100){\line(1,0){17}}
\put( 161,  96){$ 5_{4 2}$}
\put( 165, 102){\line(1,0){17}}
\put( 184,  98){$ 6_{3 3}$}
\put( 187, 109){\line(1,0){17}}
\put( 206, 105){$ 7_{2 6}$}
\put( 210, 115){\line(1,0){17}}
\put( 229, 111){$ 8_{0 8}$}
\put( 165, 122){\line(1,0){17}}
\put( 184, 118){$ 6_{4 2}$}
\put( 142, 125){\line(1,0){17}}
\put( 161, 121){$ 5_{5 1}$}
\put( 187, 126){\line(1,0){17}}
\put( 206, 122){$ 7_{3 5}$}
\put( 210, 134){\line(1,0){17}}
\put( 229, 130){$ 8_{1 7}$}
\put( 232, 143){\line(1,0){17}}
\put( 251, 139){$ 9_{1 9}$}
\put( 187, 147){\line(1,0){17}}
\put( 206, 143){$ 7_{4 4}$}
\put( 210, 147){\line(1,0){17}}
\put( 229, 143){$ 8_{2 6}$}
\put( 165, 147){\line(1,0){17}}
\put( 184, 143){$ 6_{5 1}$}
\put( 210, 158){\line(1,0){17}}
\put( 229, 154){$ 8_{3 5}$}
\put( 232, 166){\line(1,0){17}}
\put( 251, 162){$ 9_{2 8}$}
\put( 187, 172){\line(1,0){17}}
\put( 206, 168){$ 7_{5 3}$}
\put( 255, 173){\line(1,0){17}}
\put( 274, 169){$10_{0,10}$}
\put( 210, 176){\line(1,0){17}}
\put( 229, 172){$ 8_{4 4}$}
\put( 165, 178){\line(1,0){17}}
\put( 184, 174){$ 6_{6 0}$}
\put( 232, 187){\line(1,0){17}}
\put( 251, 183){$ 9_{3 7}$}
\put( 255, 199){\line(1,0){17}}
\put( 274, 195){$10_{1 9}$}
 \end{picture}

\begin{figure}[h]
\caption{\small Variation of brightness temperature $T_B$ (K) versus hydrogen
density $n_{H_2}$ for seven values of kinetic temperature $T$ (written on the 
top) for
$1_{11} - 0_{00}$ and  $2_{02} - 1_{11}$ transitions (written on the left) of TiH$_2$.
Solid line is for $\gamma = 5 \times 10^{-6}$ cm$^{-3}$ (km/s)$^{-1}$ pc, and
the dotted line for $\gamma = 5 \times 10^{-7}$ cm$^{-3}$ (km/s)$^{-1}$ pc.} \label{Fig2}
\end{figure}

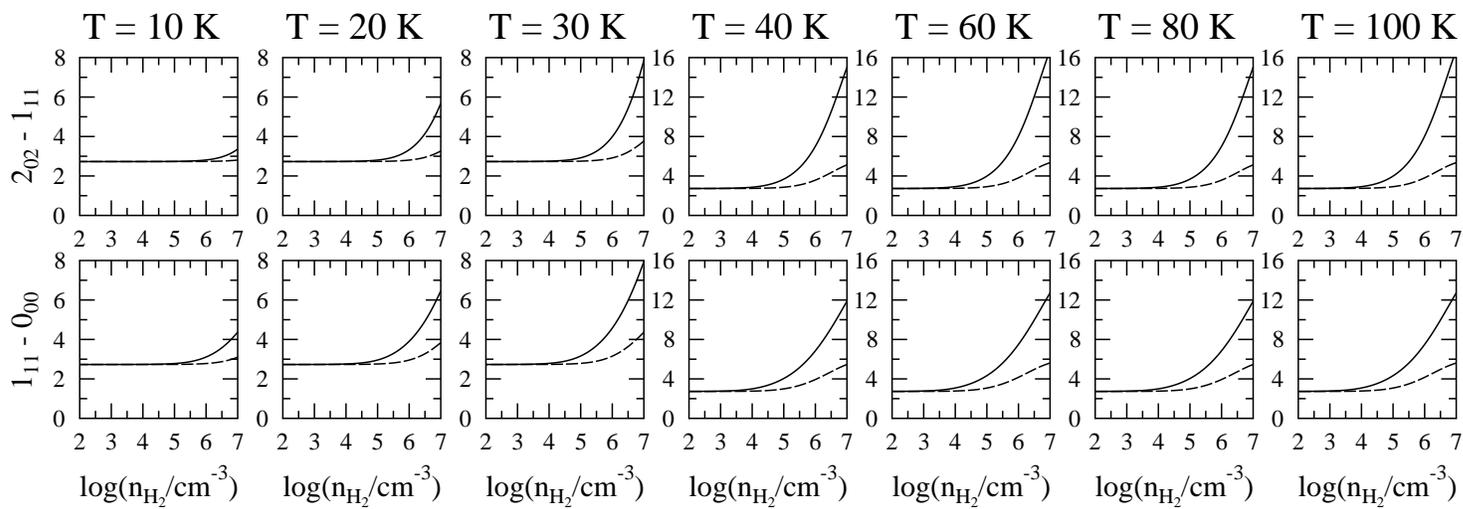
\end{document}